\documentclass[a4paper,fleqn]{cas-sc}

\usepackage[authoryear]{natbib}

\usepackage{booktabs}
\usepackage{lineno}
\usepackage{subcaption}
\usepackage{boxedminipage}
\usepackage{textpos}

\def\tsc#1{\csdef{#1}{\textsc{\lowercase{#1}}\xspace}}
\tsc{WGM}
\tsc{QE}
\tsc{EP}
\tsc{PMS}
\tsc{BEC}
\tsc{DE}

\begin{document}
\let\WriteBookmarks\relax
\def\floatpagepagefraction{1}
\def\textpagefraction{.001}
\shorttitle{Roofpedia}
\shortauthors{Authors}

\title [mode = title]{Roofpedia: Automatic mapping of green and solar roofs for an open roofscape registry and evaluation of urban sustainability}   

\author[1]{Abraham Noah Wu}[orcid=0000-0001-9586-3201]
\author[1,2]{Filip Biljecki}[orcid=0000-0002-6229-7749]
\cormark[1]
\ead{filip@nus.edu.sg}

\address[1]{Department of Architecture, National University of Singapore, Singapore}
\address[2]{Department of Real Estate, National University of Singapore, Singapore}

\cortext[cor1]{Corresponding author}

\begin{abstract}
Sustainable roofs, such as those with greenery and photovoltaic panels, contribute to the roadmap for reducing the carbon footprint of cities. However, research on sustainable urban roofscapes is rather focused on their potential and it is hindered by the scarcity of data, limiting our understanding of their current content, spatial distribution, and temporal evolution. To tackle this issue, we introduce Roofpedia, a set of three contributions: (i) automatic mapping of relevant urban roof typology from satellite imagery; (ii) an open roof registry mapping the spatial distribution and area of solar and green roofs of more than one million buildings across 17 cities; and (iii) the Roofpedia Index, a derivative of the registry, to benchmark the cities by the extent of sustainable roofscape in term of solar and green roof penetration. This project, partly inspired by its street greenery counterpart `Treepedia', is made possible by a multi-step pipeline that combines deep learning and geospatial techniques, demonstrating the feasibility of an automated methodology that generalises successfully across cities with an accuracy of detecting sustainable roofs of up to 100\% in some cities. We offer our results as an interactive map and open dataset so that our work could aid researchers, local governments, and the public to uncover the pattern of sustainable rooftops across cities, track and monitor the current use of rooftops, complement studies on their potential, evaluate the effectiveness of existing incentives, verify the use of subsidies and fulfilment of climate pledges, estimate carbon offset capacities of cities, and ultimately support better policies and strategies to increase the adoption of instruments contributing to the sustainable development of cities.
\end{abstract}


\maketitle

\begin{textblock*}{\textwidth}(0cm,-13.65cm)
\begin{center}
\begin{footnotesize}
\begin{boxedminipage}{1\textwidth}
This is the Accepted Manuscript version of an open access article published by Elsevier in the journal \emph{Landscape and Urban Planning} in 2021, which is available at: \url{https://doi.org/10.1016/j.landurbplan.2021.104167}. Cite as:
Wu AN, Biljecki F (2021): Roofpedia: Automatic mapping of green and solar roofs for an open roofscape registry and evaluation of urban sustainability. \textit{Landscape and Urban Planning}, 214: 104167.
\end{boxedminipage}
\end{footnotesize}
\end{center}
\end{textblock*}

\section{Introduction}

As urban population continues to grow, cities need to search for new spaces to feed the expansion. Traditionally, cities expanded horizontally to accommodate more housing, amenities, and public spaces, and this process took away forests and agricultural land which had a direct impact on the surrounding environment \citep{benson_2016, du2007}. According to \citet{murshed}, compact and vertical cities are considered to be more energy-efficient and environmentally friendly, and one method of increasing the compactness of cities is to build on the roofs where there is a multitude of potential uses including civic spaces, urban farms, and solar power. 

A roof serving a dual purpose is not a new invention. The flat roofs of traditional Egyptian houses were often used as an extension of their living spaces \citep{degarisdavies_1929_the}, and \citet{corbusier1927towards} believed that a roof garden is an essential element to modernist architecture. Today, luscious sky gardens with swimming pools have become a symbol of high-end residential developments, fetching considerable premiums than ordinary condominiums \citep{shukri_2017_the} and offering residents a place to get in touch with nature, which is critically correlated to happiness in an urban setting \citep{han_2019_green}.

Besides hosting civic activities, the roofscape also has enormous potential in improving the climatic and energy performance of a building and reducing or offsetting its carbon footprint, e.g.\ by providing thermal insulation, which may result in decreased energy consumption~\citep{COUTTS2013266,Jaffal:2012id,Theodosiou:2003fs,Wong.2021}. \citet{yang_2018_green} shows that green roofs can reduce 31\% of a building's heat gain on a summer day and when applied in large scale, green roofs could mitigate urban heat island (UHI) effect with an average decrease of the peak ambient temperature close to 0.9 K \citep{santamouris_2014_cooling}. These benefits would increase the thermal comfort of urban residents and reduce excessive air-conditioning energy usage due to the UHI effect \citep{WONG2003261,Wong:2003it,salamanca_2014_anthropogenic}. 
Additionally, green roofs provide undisturbed green cover in a city that could become habitats for migrating and breeding birds and can partially mitigate the loss of habitat due to increasing urbanisation \citep{partridge_2018_urban}. 

Moreover, green roofs also contribute significantly to urban stormwater management, having a mean percent rainfall retention of up to 87\% \citep{vanwoert_2005_green}, and 50 to 60\% of the rain falling on a green roof will be evaporated or transpired back to the atmosphere, hence easing the pressure on storm drainage systems \citep{jarrett_2016_green}.
Green roofs bring further environmental advantages over their unvegetated counterparts to both buildings and their surroundings~\citep{Shafique:2018hh}.
These benefits include reduction of traffic noise in urban environments~\citep{VanRenterghem:2009em,Yang:2012gx}, extending the lifetime of rooftops by shielding them from solar radiation and physical damage~\citep{Kosareo:2007fo,Carter:2008ga}, improvement of air quality and filtering air pollutants~\citep{Getter:2009dk,Yang:2008co}, improvement of aesthetic value, appearance and creation of recreational space~\citep{Villarreal:2005iy,Jungels:2013bc}, promotion of urban biodiversity~\citep{Schrader:2006jg,Brenneisen:2006vi}, and providing space for urban agriculture~\citep{Lim.2010,Whittinghill:2011fs,Beacham.2019,Palliwal.2021}. When roofs are used for urban farming, they may reduce the carbon footprint by decreasing emissions from shortened food-related transport~\citep{Goldstein:2017aa}.

Apart from green roofs, those that have a rooftop photovoltaic system (i.e. solar panels) mounted on them (in our paper, we call these \textit{solar roofs} for the sake of simplicity) are also a popular sustainable roof typology adopted by cities around the world. In comparison to green instances, solar roofs are cheaper to install and require less maintenance which makes them accessible to more homeowners. 
One key benefit of a solar roof is the financial incentive it offers. In a research by \citet{mountain_2014_australias}, Australian PV households have saved \$252 million per year between 2010 and 2012, and electricity generated from solar roofs could be half the cost than commercial providers after factoring in installation costs and government subsidies. 

It is also important to note that buildings equipped with solar roofs and building-integrated photovoltaics (BIPV) on facades can reduce operational carbon emissions of buildings by up to 50\% \citep{happle_2019_identifying}. When optimised to reach their maximum potential, solar roofs alone can help a building achieve net-zero energy consumption as seen in an increasing number of cases around the world \citep{wong_2019_nus, Sood_2019}. In contrast to the literature on green roofs, studies on the effect of solar roofs in mitigating UHI remain debatable with some of them claiming that energy absorption of solar panels could reduce UHI \citep{masson_2014_solar} while others presenting opposing views \citep{10, barron}.

Solar roofs and green roofs can be integrated as one solution that complements each other for greater environmental benefit \citep{Tablada.2018,Ciriminna.2019}. According to \citet{hui2011integration}, integrated systems could generate more electricity than standalone PV systems because green roofs improve the efficiency of the solar panels by keeping the surrounding temperature at a more suitable level for energy generation. 

Considering the benefits described above, it is not surprising that there is a multitude of studies that imply that the roofscape is a promising venue in supporting sustainable urban development \citep{HACHEM20111864, bellini2019rooftop, dubey_regenerative_2018, zamperini2014symbiotic}. Many of the studies perform large-scale geospatial simulations to estimate the potential and benefit of greening and installing PV panels on rooftops~\citep{Wong:2008hh,Santos:2016ii,RomeroRodriguez:2017ji,Bodis:2019,Joshi.2020,SHAO2020126954,Santos.2021}. For example, according to \citet{walch_2020_big}, solar panels can be installed on more than half of Switzerland's 9.6 million rooftops. The resulting power would meet more than 40\% of the Swiss electricity demand. Another study by the \citet{cityofmelbourne_2016_mapping} assesses the constraints of solar and green roof installations for existing building hoping to serve as a guide for the residents interested in upgrading to a sustainable roof typology. Recently, a crowdsourced dataset of solar panel installations in the UK was collected manually, and combined with building information from other volunteered GIS platforms such as OpenStreetMap, opening opportunities in studying the patterns of solar panel deployment \citep{stowell2020harmonised}.

However, notably scarce are studies that instead of estimating the potential, focus on the current status of the sustainable roofscape and examine the quantity and spatial distribution of current green and solar installation on rooftops at a large-scale. Such studies designed to understand the current landscape of solar and green installations in building rooftop would be useful for several purposes, including subsequent research on estimating the potential (e.g.\ by identifying roofs that are not yet utilised for either purpose), gauging the effectiveness of urban policies and verifying their mechanisms (e.g.\ subsidies by governments), tracking and monitoring environmental pledges by businesses, and estimating the current carbon offsetting capacity of cities and benchmarking them.

\begin{figure}[h]
\centering\includegraphics[width=\textwidth]{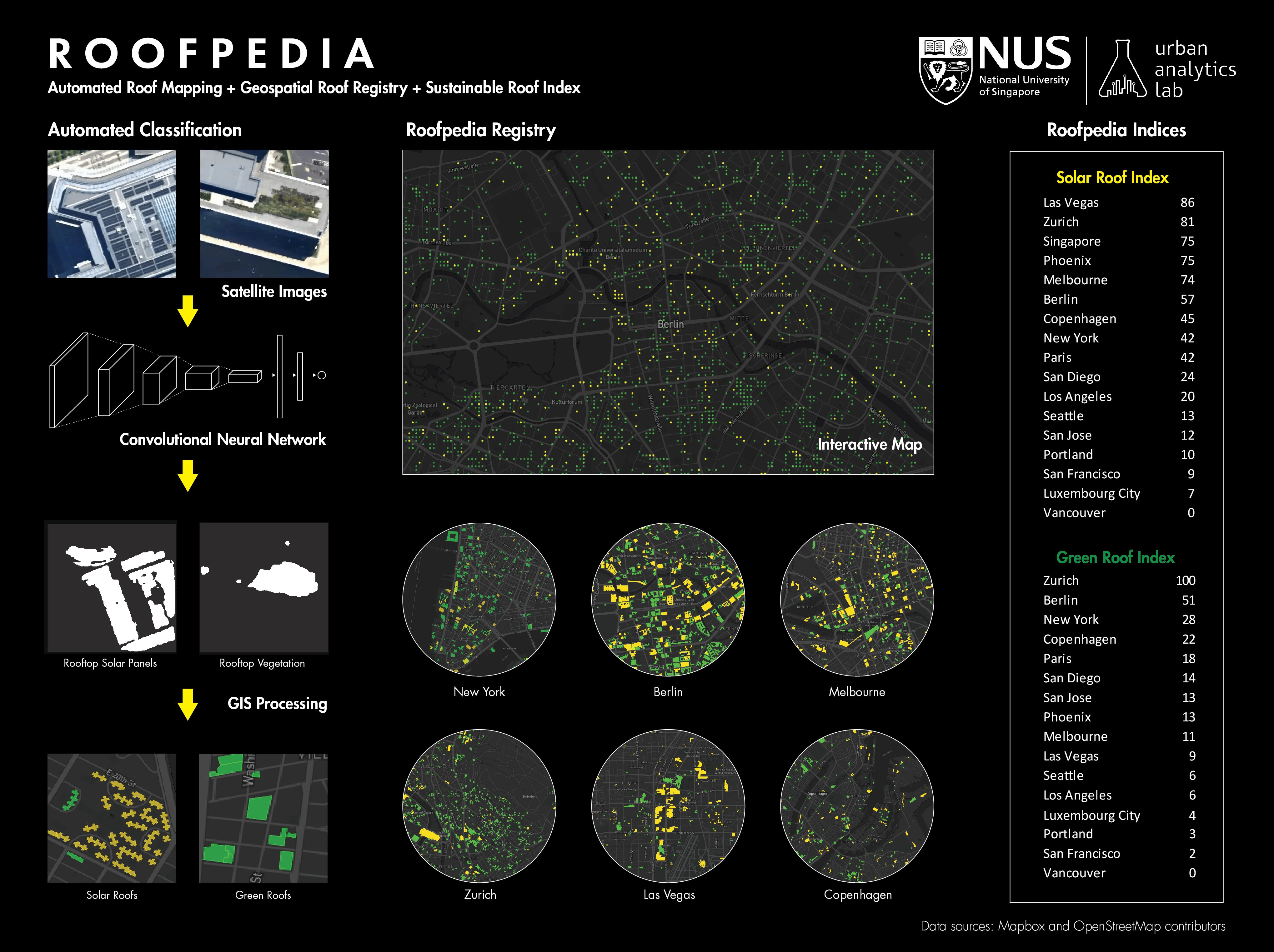}
\caption{Illustration describing the project and its organisation in three parts.}
\label{sc:mainfigure}
\end{figure}

In this paper, we bridge this gap in multiple ways, by developing a three-pronged work (Figure~\ref{sc:mainfigure}). First, we develop an automated method to identify rooftops that are vegetated and/or have photovoltaic system installations. The method relies on high-resolution satellite imagery is available for an increasing number of cities around the world, facilitating replication and scalability of our work. Second, we implement the method on 17 cities to create a \textit{cadastre} of the roofscape, revealing the quantity and spatial distribution of rooftops that contribute towards a city's sustainable development. Our dataset contains information on more than one million buildings, and we are making this sustainable roofscape inventory available publicly, aiding researchers, practitioners, local governments, and the public to understand the current status of the roofscape in the context of sustainable urban development and achieving carbon neutrality. Third, our registry enables us to benchmark and compare cities with regard to the utilisation of their roofscape.
We establish the level of penetration of green and solar roofs in each city, and quantify it, creating the Roofpedia Index. The scores for the 17 cities in our study are computed, and the cities are compared.

Some of the aspects of the project and its name --- Roofpedia --- are inspired by Treepedia \citep{seiferling2017green, senseablecitylab}, a project to measure and map the amount of street greenery in cities from the pedestrian perspective, and compare cities around the world. A notable characteristic of our modular work is that it enables replication and extensibility, thus, additional cities can be included in Roofpedia in the future --- from mapping the roofscape to calculating their index for benchmarking purposes, and potentially supporting future extensions to include other typologies of rooftops such as social uses.
Furthermore, this service may potentially provide a basis for a crowdsourced platform that would integrate additional information on rooftops.

The paper is organised as follows.
In Section~\ref{sc:background} we expand on the introduction and continue the literature review.
Section~\ref{sc:mapping} describes our methodology for automatically mapping green and solar roofs, and evaluates the results we achieved.
Section~\ref{sc:index} analyses the results, presents our database, and the derived Roofpedia Index to assess the performance of cities.
In Section~\ref{sc:discussion} we discuss the work, expose limitations, and map out ideas for future work.
Section~\ref{sc:conclusion} concludes the paper.

\section{Background}\label{sc:background}
\subsection{Sustainable roof policies and documentation}\label{sc:policies}

From the previous section on the benefits of a sustainable roofscape, it is evident that the roof could offer much more to a city rather than just been a plain enclosure for shelter and interior comfort. The roofscape of a city can contribute positively to a city's environment and quality of life, especially when instruments for sustainable development such as vegetation and PV panels are applied widely. 

The potential is even greater when considering that not only new buildings may be enriched with green and solar installations. For example, many existing unvegetated roofs can be converted into green roofs~\citep{Wilkinson:2013ey}, which may be especially appropriate for buildings with flat roofs and sufficient structural strength~\citep{Clark:2008ke,Stovin:2010iy,Castleton:2010cb}.

While green roofs are a viable expenditure from the societal, environmental and economical perspective, they come at a cost, and their installation is inhibited by high investments~\citep{Nelms:2007bq,Bianchini:2012cl,Sproul:2014cb}.
To offset these costs, many local governments worldwide have introduced various incentives and guidelines to encourage the utilisation of rooftops spaces for activities that contribute to sustainable development~\citep{Claus:2012gs,Chen:2013bg}.
For example, sky gardens for communal uses are exempted from Gross Floor Area calculations in Hong Kong \citep{hongkongbuildingsdepartment_2019_green}, and Singapore's National Parks Board will fund up to 50\% of installation costs of green roofs for developers \citep{nationalparksboard_2009_skyrise}.
The incentives in these two land-scarce and densely populated jurisdictions also suggest the proliferation of the role of roofscapes in transcending the frontiers of space utilisation, especially in promoting carbon offsetting. In the case of solar roofs, private PV panel owners in the United States are allowed to `turn-back' their meter if they use less energy than what the solar panel generates \citep{usdepartmentofenergy_2020_gridconnected}. Residents in the Australian state of Victoria can enjoy a rebate for the installation of a solar panel system with a four-year, interest-free loan for the remaining amount \citep{solarvictoria_2020_solar}. While the above incentives could potentially increase the adoption of sustainable roofs, publicly available data on the current state of adoption in most cities is very limited, making it difficult to assess the effectiveness of current policies or to understand the current state of adoption. 

Some projects have endeavoured to tackle this issue by mapping rooftop typologies at the city-scale. There are two initiatives that are closely related to our work. First, the \citet{cityofmelbourne_2016_mapping} created a one-of-a-kind roof registry that maps the city's existing green and solar roofs and also potential roofs that could be converted to sustainable roofs. Another study was done in Berlin by the \citet{senate_2017_green}. An automatic mapping was carried out with GIS processing that returned detailed statistical data on the area and location of green roofs in the city. Both projects offer an interactive map of the dataset on sustainable roofs and provide an open database for further research. 

These two projects set a commendable example in popularising roof data as they enable everyone to explore the roofscape intuitively and potentially sparking interest in the topic. However, these projects have their fair share of limitations. For example, the project in Berlin is restricted to mapping green roofs. Further, both are focused on a single city, not allowing easy comparisons on how a city fares in comparison with others when it comes to sustainable installations on their rooftops. Additionally, no statistical validation was carried out to assess the accuracy of their methodology, nor the work has been published in international scientific outlets.

In this paper, we seek to overcome their shortcomings. We advance the steps taken by these two projects further and aim to map both green and solar roofs across multiple cities to create a global registry of sustainable roofscapes. We aim to assess the feasibility and accuracy of an automated mapping methodology and present the information intuitively to the public. It is important to underline that our work is not confined to a single city, as our cross-city analysis not only includes more than a dozen cities, but it is further scalable as long as satellite imagery of comparable quality as the ones in the training set is available. Such scalability has allowed us to create a cross-city benchmark (Roofpedia Index) that measures the penetration of solar and green roofs across 17 different cities. Furthermore, we have provided open access to the Roofpedia prediction pipeline where users could carry out the mapping and estimations using their own imagery. Additionally, our methodology of using satellite imagery allows us to examine the temporal evolution of sustainable rooftops, which would enable multiple use cases in the future, e.g.\ understanding the effectiveness of policies on a temporally fine scale.
As photovoltaic capacity is dynamic and rapidly evolving, it is important to track changes through time.

\subsection{Automated roof mapping with deep learning}

In recent years, deep learning, especially Convolutional Neural Networks (CNN) \citep{7478072} offer new opportunities for object detection in optical remote sensing images \citep{CHENG201611} and it has been used for various applications in the built environment \citep{Middel.2019,Chen.2021r5p}. \citet{basu_ganguly_mukhopadhyay_dibiano_karki_nemani_2015} demonstrated that deep learning algorithms can classify satellite image tiles to an accuracy of 97.95\%. Going further in granularity, \citet{6778050} created a pipeline for vehicle detection in satellite images using Deep Convolutional Network.

There exists a multitude of research on the feasibility of mapping the roofscape with CNNs with satellite or aerial images. \citet{chen2019deep} showed that CNNs can detect and segment building boundaries of small detached houses from aerial images in the city of Christchurch. \citet{Castello_2019} demonstrated a deep learning method that predicts solar panels at pixel level with an accuracy of about 94\% and an Intersection over Union of up to 64\% using high-resolution aerial photos, which is a significant leap compared to traditional machine learning approaches \citep{MALOF2016229}. However, the research was only an evaluation of a small region, and its scalability remains to be tested.

While aerial images are still limited to selected areas, satellite images have a wider coverage, and previous studies had demonstrated that deep learning methods on satellite images can still provide reliable predictions of the shapes and the sizes of building roofs or building footprints \citep{nosrati_saeedi_2009, lee2019}. Therefore, deep learning performed on satellite images will not be limited to a specific area but has the potential to scale up to cover myriads of cities around the globe.

Using a CNN architecture called Inception-v3 and a large-scale training set of 366,467 images, \citet{yu_2018_deepsolar} created a solar installation database for the contiguous U.S. This project demonstrates the scalability of computer vision algorithms in mapping solar panels from satellite images, and their results are encouraging for the development of our project. However, the study provides visualisation only at the scale of census tracts, unlike the Melbourne study where the resolution is at the building level. It is also difficult to use the same method on green roofs as there is yet to be a large labelled green roof dataset available. Further, while appreciating the scope of the study given the large extent of the U.S., the project is still focusing on only one country.

Nevertheless, compared to traditional spectral classification, CNNs can be used on lower quality images that are widely available in the public realm. Using a modified CNN pipeline with U-Net architecture, in the next section, we present an automated method to identify both green and solar roofs from satellite images and assign corresponding labels to building datasets derived from OpenStreetMap to create a roof registry at the building level for spatial visualisation and further analyses. 

Besides the novelties and contributions discussed in Section~\ref{sc:policies}, our method has several advantages compared to existing methods mentioned in this section. Firstly, we present work at a substantially higher spatial resolution than the current state of the art (i.e.\ census tract level to building level). Secondly, the model is created with international scalability in mind and it can be used in multiple cities around the world with different urban morphology and building typology. Thirdly, our method identifies both green and solar roofs, which adds another dimension to understanding the sustainability of the roofscape across cities. Finally, as the cities are assessed with the same model and within the same system, comparisons between cities would be fairer and easier than comparing results from different research projects. This advantage enables us to create a sustainable index that ascribes standardised scores to the state of sustainable roof practice in each city.
Therefore, one of the outcomes of the work is not only a database of such roofs across many cities, but also gauging how sustainable and effective cities are when it comes to unlocking the space provided by rooftops.

\section{Automated roof classification}\label{sc:mapping}
\subsection{Data and study areas}

To demonstrate the scalability and feasibility of our methodology, 17 morphologically diverse and geographically distributed cities spanning Europe, North America, Australia, and Asia are selected as our study areas. These are: Berlin, Copenhagen, Las Vegas, Los Angeles, Luxembourg City, Melbourne, New York, Paris, Phoenix, Portland, San Diego, San Francisco, San Jose, Seattle, Singapore, Vancouver, and Zurich.

The satellite imagery used in this study is retrieved with the Mapbox static tiles API, except for Luxembourg for which the data is acquired from the open data by the Luxembourgish Land Registry and Topography Administration, affirming that our method is not constrained to one data source.
Mapbox has a generous free tier usage limit enabling obtaining imagery of several cities. According to Mapbox, the images available for access are from a combination of different sources, including Maxar’s Vivid products for much of the world, Nearmap and USDA’s NAIP 2011–2013 in the contiguous United States, and open aerial imagery from Denmark, Finland, Germany, and other regions. These datasets offer resolutions of about 50 cm and higher in major cities in the world, and would be an effective way to assess the scalability and adaptability of our methodology. Data on building footprints have been extracted from OpenStreetMap, which is being increasingly used in research \citep{Cerri.2020,Feldmeyer.2021}, and which has a very good level of completeness for the study areas we have selected~\citep{Fan:2014kz,Biljecki.2020}. The building footprints serve a dual purpose: they filter out greenery and solar panels not installed on rooftops, and they are used to integrate the results of our work, enriching the building data. The visualisations of the final result are rendered on Mapbox for public access and both the dataset and code are available openly on Github.

In our research, we have included a few more cities besides the listed ones (e.g.\ Washington D.C. and Marseille). 
However, eventually we did not proceed forward with including them in our database and index, due to a deviating performance of the model to identify sustainable roofs, which was largely caused by external factors such as quality of imagery and presence of particular features in these cities (e.g.\ a large density of skylights that tend to be misidentified as solar panels).
Nevertheless, considering these cities was vital for understanding the data requirements and exposing limitations of the method, and further examples are given in Section~\ref{sc:results}.

\subsection{Methodology}

\begin{figure}[h]
\centering\includegraphics[width=1\linewidth]{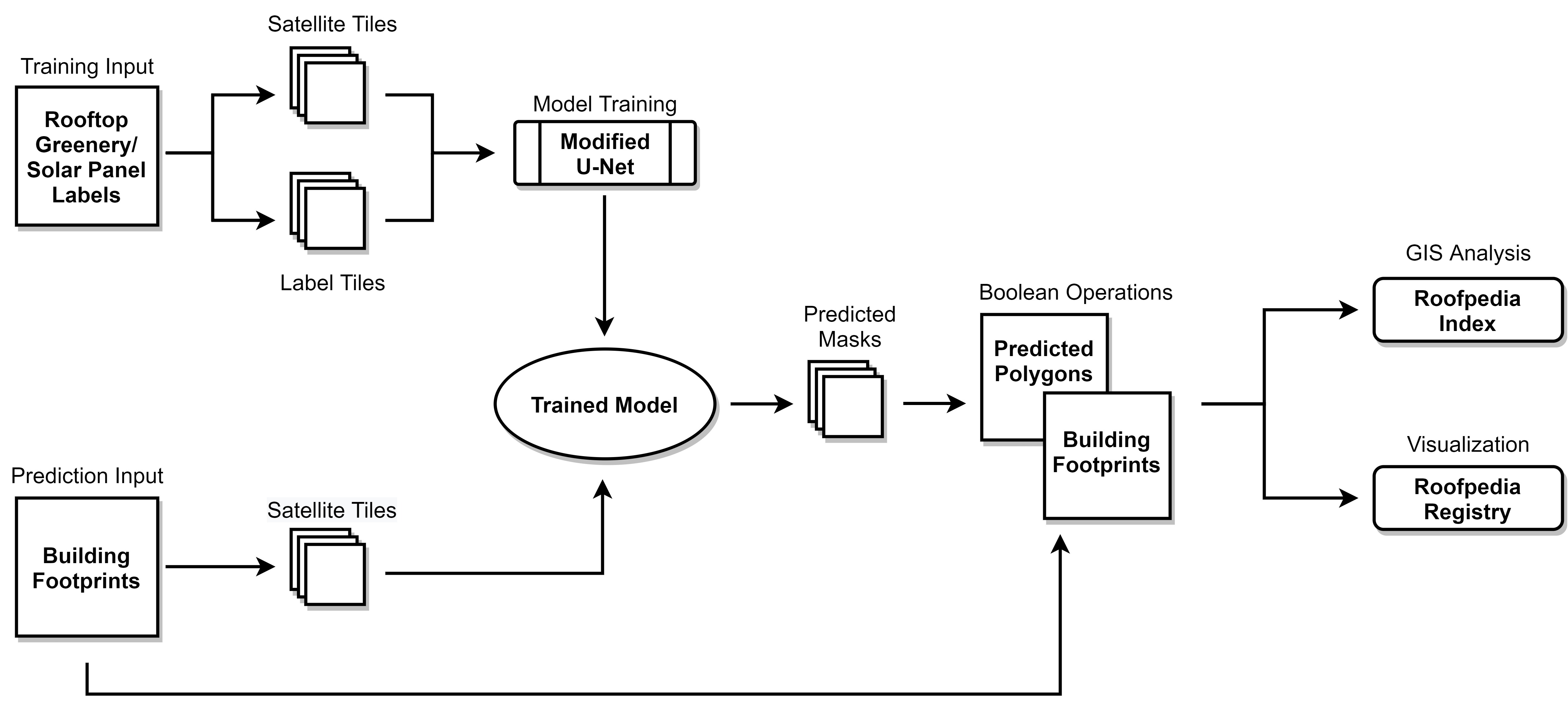}
\caption{Overview of the methodology of Roofpedia.}
\label{fig:method}
\end{figure}

Our methodology can be broken down into four steps (Figure~\ref{fig:method}). 
First, we label the training images across multiple diverse locations and tile the satellite images to a uniform shape for the deep learning algorithm. The polygonal labels required for the training are manually labelled across multiple cities to create a wide range of examples in different urban contexts and image conditions.
Secondly, the processed dataset is trained in a Convolutional Neural Network based on a modified U-net architecture. The network is initialised with a pre-trained model (ResNet50) to improve the accuracy of the prediction. 
Thirdly, the building footprints of the predicted areas are passed in as new inputs and the probabilities predicted by the CNN are converted into georeferenced polygons to tag the respective building footprints with either `Solar' or `Green' or both.
Lastly, the resulting, semantically enriched building footprints are analysed quantitatively and visualised spatially to create the Roofpedia Registry and the Roofpedia Index.

\subsubsection{Labelling and tiling}

\begin{figure}[h]
\centering\includegraphics[width=0.75\linewidth]{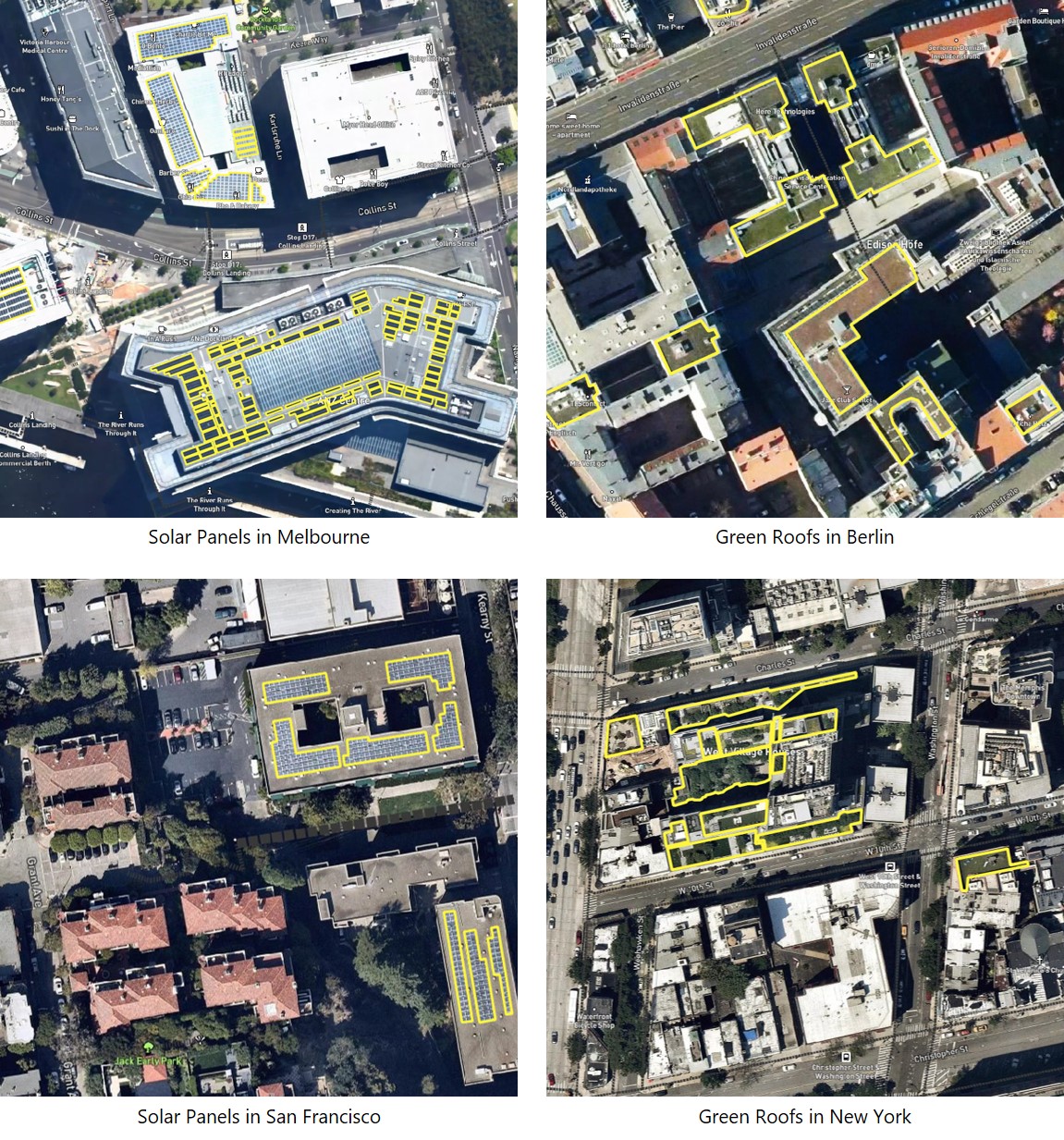}
\caption{Labelling solar panels and rooftop greenery across cities.}
\label{fig:labelling}
\end{figure}

Figure~\ref{fig:labelling} shows a sample polygon label for rooftop solar panels and rooftop greenery in multiple cities. In total, 1812 polygons have been labelled across eight cities to capture a wide range of rooftop greenery and solar panels. Specifically, there are 545 unique labelled rooftop greenery summing up to an estimated 208,116 square metres in area and 1,254 solar panels summing up to an estimated 175,698 square metres in area. Converted into individual tiles for training, validation and testing, there are a total of 1,517 labelled tiles for rooftop greenery and 1,380 labelled tiles for solar panels.

For training, we have used the imagery of 8 instead of all 17 cities in our study to determine whether the trained models will be scalable in new locations.
In each of the cities, areas are randomly selected during labelling and thus the labelled data do not represent all green and solar roofs in the city. Nevertheless, they are sufficiently representative of the green and solar roof typology in the city, and as the continuation of this paper will demonstrate, the model trained on these locations scales successfully in new areas without training data, facilitating the extension of the work to easily and accurately include new cities after the publication of this paper.

\begin{figure}[h]
\centering\includegraphics[width=0.75\linewidth]{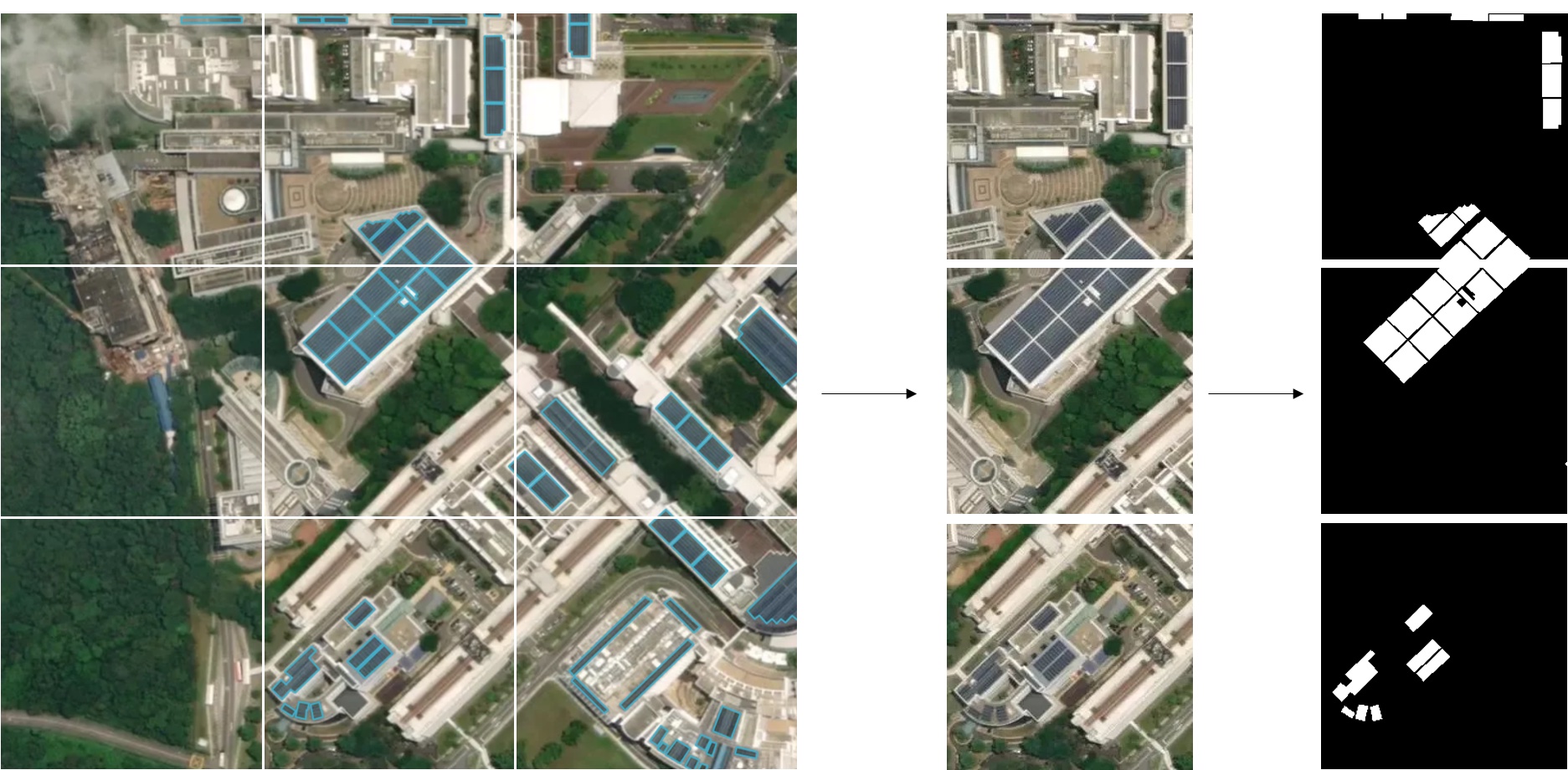}
\caption{Tiling of satellite images and labels.}
\label{fig:tiling}
\end{figure}

After creating training labels, we slice the tiles and the vector polygon masks of the satellite images using a zoom level of 19 according to the Slippy Map tiling convention described by OpenStreetMap. This zoom level enables the model to focus on roof objects and minimise confusion with objects outside of the building boundaries.
The images and masks are further sliced into 256~x~256-pixel datasets for training. Labelled polygons are similarly sliced and converted into image masks for training  (Figure~\ref{fig:tiling}). In total, 2,897 images are prepared with 80\% serving as training data and 20\% as test data to validate the accuracy of the model.

\subsubsection{Image segmentation with deep learning}

\begin{figure}[tbp]
\centering\includegraphics[width=0.75\linewidth]{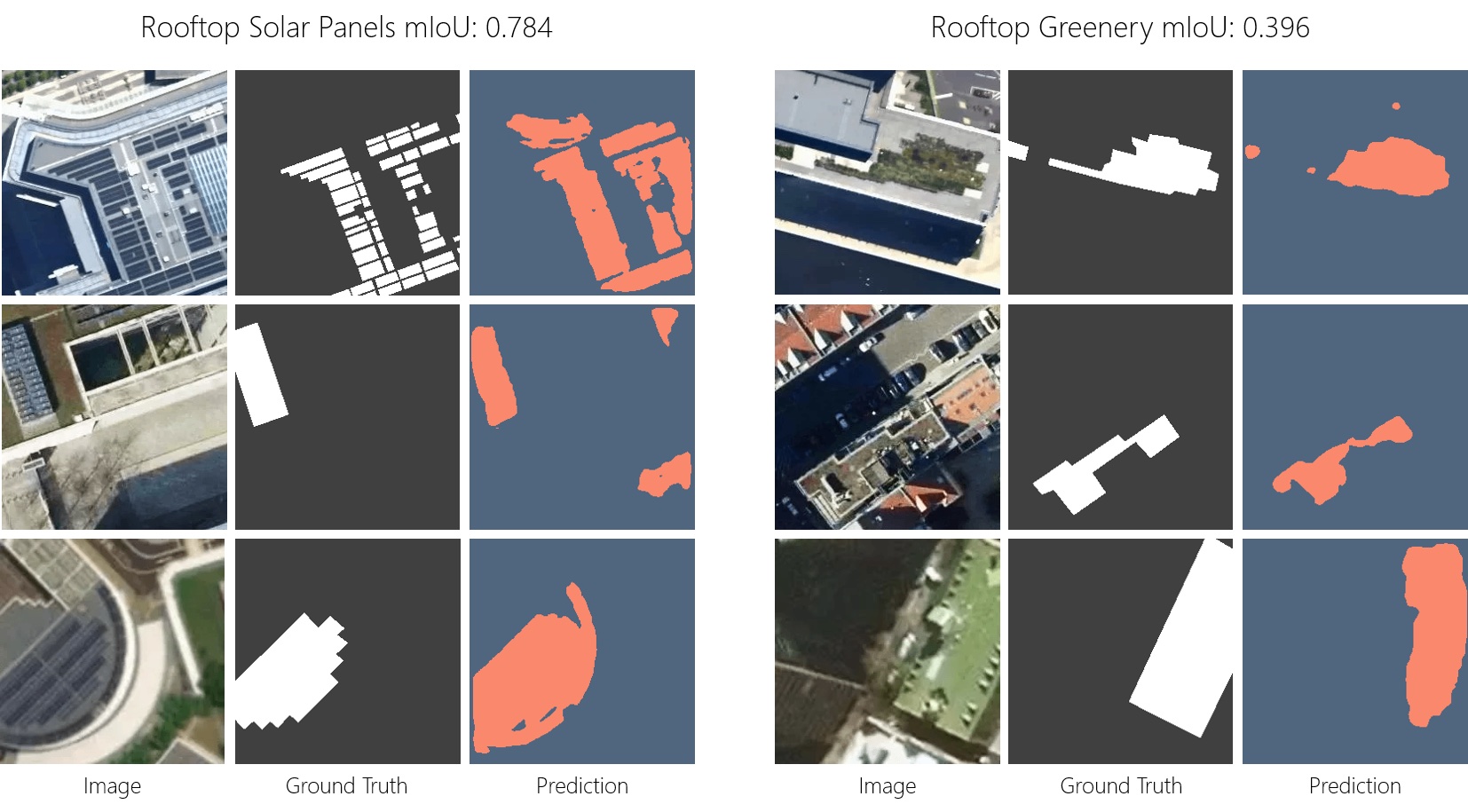}
\caption{Model prediction results compared with ground truth.}
\label{fig:presults}
\end{figure}

The deep learning model used is a type of Fully Convolutional Neural Network \citep{7478072}, which excels at image boundary identification and instance segmentation. The model is modified from U-Net \citep{ronneberger2015u} by replacing the encoding section of the U-Net with a pre-trained Resnet50 network \citep{he2015deep}. 
The U-Net architecture is chosen because of its optimisation on augmented data which uses available training samples more efficiently. As such, a U-Net can be trained from very few images and still provide reliable results. By replacing the pre-trained Resnet50 as the encoding layer for the U-Net, we leverage the power of transfer learning to reduce false negatives \citep{shi2019deep}, and further improve the validation accuracy of the model \citep{shin2016deep, hussain2018study}.

We use PyTorch \citep{NEURIPS2019_9015} for building the deep learning model and we have borrowed a preprocessing pipeline from Robosat, a satellite feature extraction toolbox \citep{ng_2018_scalable}. We created our own pipeline integrating these packages, which is available publicly for reproducibility.

To avoid confusion in shared features between greenery and solar panels, the model is trained on the solar dataset and green dataset separately, learning the weights independently for each feature. During training, the model learns the respective features that make up a patch rooftop greenery or a solar panel from the input image and mask and validates its accuracy against the test set. 

Figure~\ref{fig:presults} shows the results of the prediction model against the ground truth in the evaluation set for the green and solar models respectively. The performance of the model is measured by Mean Intersection over Union (mIoU) which measures the number of pixels common between the ground truth and prediction masks divided by the total number of pixels present across both masks. The model for identifying solar portions of roofs achieved a mIoU score of 0.784 while the green roof counterpart achieved a mIoU of 0.396. Compared to solar panels which have a well defined edge, some green roofs contain trees and vegetation that do not have a defined edge. This fuzziness has resulted in lower mIoU score for the green roof model as it struggles to find these edges. Nevertheless, this shortcoming does not have a major influence over the quality of the final result as only the locations of the predicted polygons are used in GIS operations.

\begin{figure}[h]
\centering\includegraphics[width=0.8\linewidth]{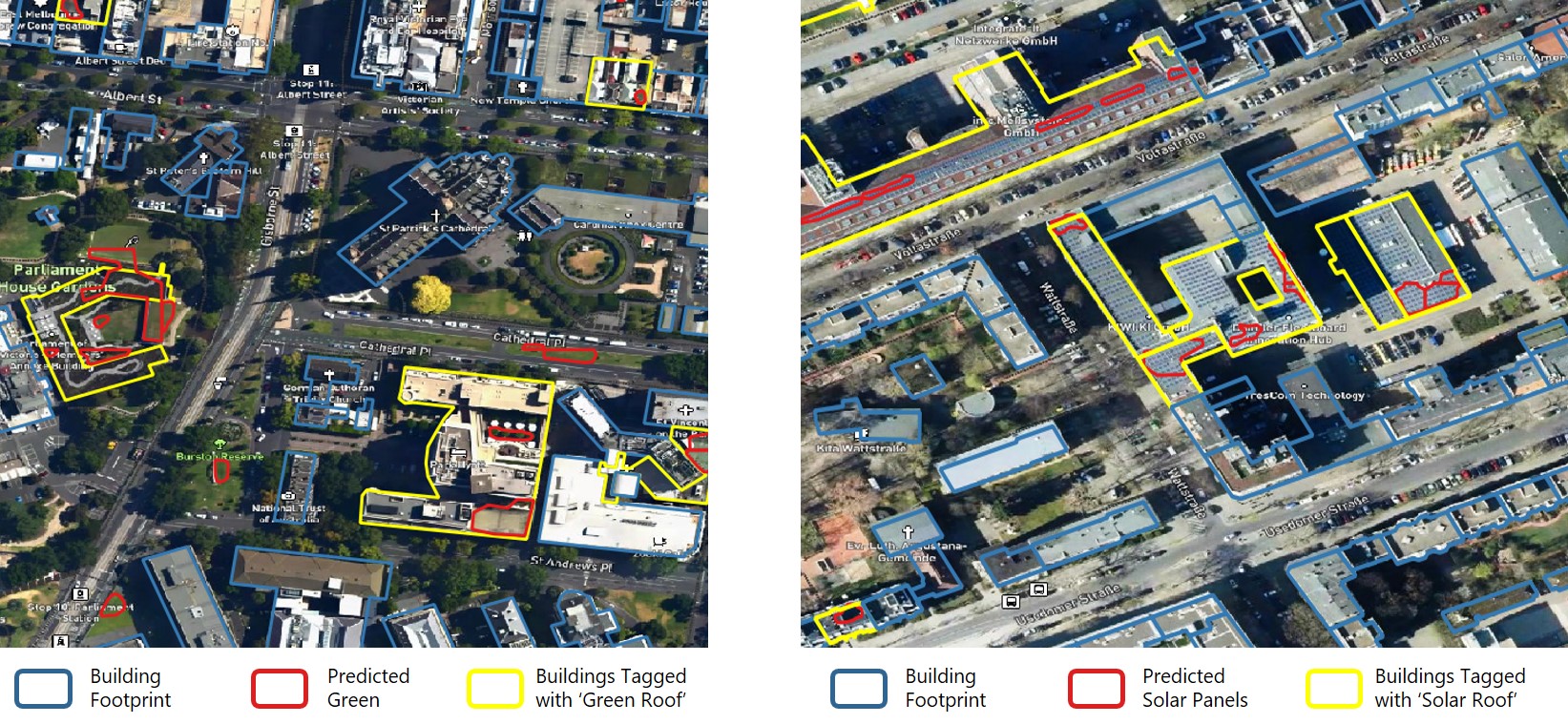}
\caption{Geospatial engineering of the predicted result: tagging buildings with sustainable rooftops based on detected greenery and solar panels on them.}
\label{fig:distribution}
\end{figure}

\subsubsection{Classification with geospatial engineering}

Even though the training labels of the deep learning model only contain solar panels and greenery on building rooftops, the entire image tile is passed through the model during training and inference. This process causes features that look like solar panels or greenery outside of the building rooftops to be picked up. For example, green patches in a park or trees on the streets could potentially be identified since they share similar features as rooftop gardens. Open-air carparks also have a similar grid-like feature that could be confused as solar panels.
While this issue can be technically solved by complex image pre-processing procedures within the model architecture (i.e.\ creating a cropped image for each building), it will increase the size of the data and the complexity of the model, and thus increase the computational complexity for training and prediction. 

To avoid this issue, we added a post-processing algorithm to remove false positives and restrict the predictions only within the building polygons. The prediction results from the deep learning model are converted into georeferenced polygons for GIS processing. When converting the pixel masks into predicted polygons for GIS operations, there is a small degree of rectification to simplify the polygons to reduce the number of edges while maintaining the general shape of the prediction. In addition, a denoiser is added to remove `speckles' to reduce the number of False Positives. The setting of the denoiser is fine-tuned to provide the best balance between accuracy and precision.

Following the conversion, the resultant polygons that do not intersect building polygons are removed, and polygons that intersect building polygons but are too irregular or insignificant are ignored since we have noticed that these are predominantly false positives. Finally, building footprints that contain or intersect with the cleaned prediction set are assigned a label depending on whether they have a solar or green roof, or both (Figure~\ref{fig:distribution}).

\subsection{Results and evaluation}\label{sc:results}

To determine the scalability and accuracy of the prediction, 11 neighbourhoods in different cities are selected for evaluation. These areas are not included in the original training set and their ground truths are labelled manually. As such, the results obtained is unbiased and effective in measuring the adaptability of the model across cities. Furthermore, some of the neighbourhoods are in cities excluded from training data which will help to measure the adaptability of the model when applied to a new city where it has not been trained. The size of the neighbourhoods varies with a range of 475 to 3425 buildings, and all buildings in the selected areas are carefully labelled as the ground truth, which is then compared against the predicted results.

\subsubsection{Metrics for evaluation}

We evaluate the performance of the methodology in both the number of buildings (Count) and the area of rooftops (Area) identified with green or solar roofs. A model that predicts Count accurately allows us to analyse the distribution of the identified rooftops in the urban fabric whereas a model that predicts Area accurately allows us to analyse how extensive a city has adopted sustainable roof typologies amongst its total roof area. Learning from the performance metrics proposed by \citet{7120453}, the following variables are used for calculating the performance of the method . These variables are also shown as columns in Tables~\ref{tab:green-count}--\ref{tab:solar-area}.

\begin{itemize}
\item Total Count/Area (Total): Total number/area of building footprints in the study region
\item Truth: Total number/area of building footprints in the Ground Truth set
\item Prediction (TP + FP): Total predicted number/area of building footprints predicted by the model
\item True Positive (TP): Both predicted and ground truth data classify roof as green or solar.
\item False Positive (FP): Only the predicted data classify roof as green or solar.
\end{itemize}

Accuracy of the prediction is measured by the Percentage of True Positives (\%Matching) in terms of Count or Area. This value is obtained by dividing the Count or Area of the True Positives against the Total Count or Area of the Ground Truth in Equation~\ref{eqn1}.
Precision is measured by the Percentage of False Positives (\%FP) in terms of Count or Area. This value is obtained by dividing the Count or Area of the False Positives (FP) against the Count or Area of all buildings (Total) in the dataset in Equation~\ref{eqn2}.
The proportion of roofs that contains greenery or solar panels both in terms of Count or Area is represented by \%Cover. This metric helps to interpret the extensiveness of the coverage of solar and green roofs in the regions according to the true positives in the prediction in Equation~\ref{eqn3}.

\begin{align}
\label{eqn1}
& \textrm{\%Matching} = 100 \times (\textrm{TP}/\textrm{Truth})\\
\label{eqn2}
& \textrm{\%FP} = 100 \times (\textrm{FP}/\textrm{Total})\\
\label{eqn3}
& \textrm{\%Cover} = 100 \times (\textrm{Matching (TP)}/\textrm{Total})
\end{align}

\subsubsection{Green roof results}

Table~\ref{tab:green-count} and Table~\ref{tab:green-area} present the prediction results of green roofs against the manually labelled ground truth. Using absolute Count as the metric, the accuracy of the model reached 77.45\% of the ground truth, and the average \%FP reached 1.92\%. Comparing the Area of the prediction with the ground truth, accuracy increased to 87.59\% but \%FP also increased, to 3.66\%. 

\begin{table}[tbp]
  \centering
  \scriptsize
  \caption{Green roof result by Count (building classification).}
    \begin{tabular}{lrrrrrrrr}
    \toprule
    \textbf{Region} & \textbf{Area} & \textbf{Bldg.} & \textbf{Truth} & \textbf{Pred.} & \textbf{Matching} & \textbf{\%Matching} & \textbf{\%FP} & \textbf{\%Cover} \\
    & \textbf{(km$^{2}$)} & \textbf{Count} & & & & & & \\
    \midrule
    Berlin 1 & 1.20  & 754   & 153   & 158   & 121   & 79.08 & 4.91 & 16.05 \\
    Berlin 2 & 1.21  & 1009  & 93    & 108   & 86    & 92.47 & 2.18 & 8.52 \\
    Melbourne & 4.50  & 1079  & 26    & 44    & 24    & 92.31 & 1.85 & 2.22 \\
    New York 1 & 1.02  & 757   & 145   & 89    & 81    & 55.86 & 1.06 & 10.70 \\
    New York 2 & 2.05  & 1481  & 73    & 56    & 52    & 71.23 & 0.27 & 3.51 \\
    Paris & 1.70  & 3425  & 83    & 78    & 64    & 77.11 & 0.41 & 1.87 \\
    Seattle & 1.80  & 598   & 35    & 32    & 29    & 82.86 & 0.50 & 4.85 \\
    Zurich & 1.13  & 475   & 115   & 99    & 79    & 68.70 & 4.21 & 16.63 \\
    \midrule
    \textbf{Average} &       &       &       &       &       & \textbf{77.45} & \textbf{1.92} & \textbf{8.04} \\
    \bottomrule
    \end{tabular}%
  \label{tab:green-count}%
\end{table}%

\begin{table}[h]
  \centering
  \scriptsize
  \caption{Green roof result by Area.}
    \begin{tabular}{lrrrrrrrr}
    \toprule
    \textbf{Region} & \textbf{Area} & \textbf{Bldg.} & \textbf{Truth} & \textbf{Pred.} & \textbf{Matching} & \textbf{\%Matching} & \textbf{\%FP} & \textbf{\%Cover} \\
    & \textbf{(km$^{2}$)} & \textbf{Area (m$^{2}$)} & \textbf{(m$^{2}$)}& \textbf{(m$^{2}$)}& \textbf{(m$^{2}$)} & & & \\
    \midrule
    Berlin 1 & 1.20   & 673152 & 234785 & 234616 & 180603 & 76.92 & 8.02 & 26.83 \\
    Berlin 2 & 1.21  & 533421 & 76766 & 113235 & 73418 & 95.64 & 7.46 & 13.76 \\
    Melbourne & 4.50   & 1042319 & 61436 & 108072 & 60071 & 97.78 & 4.61 & 5.76 \\
    New York 1 & 1.02  & 364340 & 123435 & 101075 & 91310 & 73.97 & 2.68 & 25.06 \\
    New York 2 & 2.05  & 889578 & 263599 & 245665 & 245431 & 93.11 & 0.03 & 27.59 \\
    Paris & 1.70   & 866545 & 91143 & 104551 & 79547 & 87.28 & 2.89 & 9.18 \\
    Seattle & 1.80   & 771931 & 76402 & 70469 & 68408 & 89.54 & 0.27 & 8.86 \\
    Zurich & 1.13  & 221836 & 97946 & 92136 & 84677 & 86.45 & 3.36 & 38.17 \\
    \midrule
    \textbf{Average} &       &       &       &       &       & \textbf{87.59} & \textbf{3.66} & \textbf{19.40} \\
    \bottomrule
    \end{tabular}%
  \label{tab:green-area}%
\end{table}%

In both cases, \%FPs are under 5\% for most of the areas except Berlin 1 (Stadtmitte) and Berlin 2 (Friedrichshain). The percentage of matching (\%Matching) is generally above 80\% in terms of Area except for New York 1 (Hudson Square) and above 70\% in terms of Count except for Zurich (Alstetten) and New York City 1 (Hudson Square).

\begin{figure}[tbp]
\centering\includegraphics[width=0.95\linewidth]{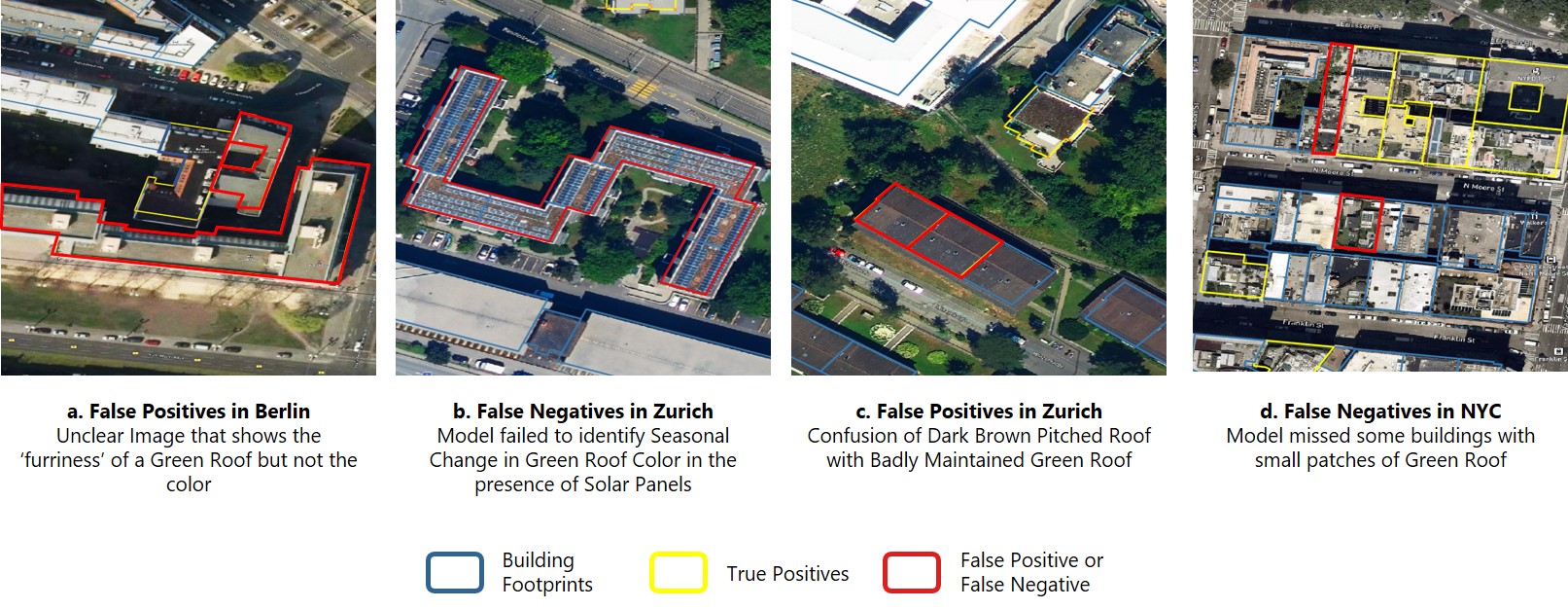}
\caption{Examples of green roof model confusions.}
\label{fig:green-confusions}
\end{figure}

Looking closely into these regions, we observed that some degree of confusion had occurred for the model. As shown in Figure~\ref{fig:green-confusions}, some roofs in Berlin (7a), have a light green and `furry' appearance which the model confuses as a green roof. In the case of Zurich (7b, 7c), the model failed to recognise seasonal changes in green roof colour (browning of leaves in autumn) in the presence of solar panels. It successfully picked up some badly maintained green roofs (dark brown in colour with a furry texture) but confused it with dark brown pitches roofs in the vicinity. Finally, the model also failed to identify some green roofs in smaller buildings such as in New York City (7d). This issue might be caused by the removal of noise in the prediction pipeline where some true positives are sacrificed to reduce the percentage of false positives (in the case of New York City, removing noise contributed to an average 67.6\% reduction of false positives (4.17\% to 1.35\%) at the cost of 5.48\% decrease of accuracy (88.38\% to 83.54\%)).

\subsubsection{Solar roof results}

Table~\ref{tab:solar-count} and Table~\ref{tab:solar-area} present the prediction results of solar roofs against the manually labelled ground truth. Using the same method of calculation as that of the green roofs, the accuracy of the solar roof model in terms of Count reached 91.90\% and an average \%FP of 1.00\%. Accuracy in terms of calculating the Area increased to 94.06\% while \%FP also increased to 4.42\%. This performance suggests that larger buildings might have higher chances of having features that confuse the model into categorising it as a solar roof building.

\begin{table}[h]
  \centering
  \scriptsize
  \caption{Solar roof result by Count (building classification).}
    \begin{tabular}{lrrrrrrrr}
    \toprule
    \textbf{Region} & \textbf{Area} & \textbf{Bldg.} & \textbf{Truth} & \textbf{Pred.} & \textbf{Matching} & \textbf{\%Matching} & \textbf{\%FP} & \textbf{\%Cover} \\
    & \textbf{(km$^{2}$)} & \textbf{Count} & & &  & &  &\\
    \midrule
    Berlin 1 & 2.31   & 2251  & 87    & 90    & 72    & 82.76 & 0.80 & 3.20 \\
    Berlin 2 & 1.22   & 754   & 31    & 43    & 28    & 90.32 & 1.99 & 3.71 \\
    Copenhagen & 1.64   & 1397  & 9     & 12    & 7     & 77.78 & 0.36 & 0.50 \\
    Marseille & 3.51   & 5639  & 73    & 79    & 72    & 98.63 & 0.12 & 1.28 \\
    Melbourne & 4.52   & 1079  & 36    & 59    & 36    & 100.00 & 2.13 & 3.34 \\
    New York 1 & 1.54   & 1414  & 26    & 26    & 26    & 100.00 & 0.00 & 1.84 \\
    New York 2 & 5.52   & 3079  & 28    & 27    & 24    & 85.71 & 0.10 & 0.78 \\
    Washington D.C. & 2.10   & 562   & 37    & 51    & 37    & 100.00 & 2.49 & 6.58 \\
    \midrule
    \textbf{Average} &       &       &       &       &       & \textbf{91.90} & \textbf{1.00} & \textbf{2.65} \\
    \bottomrule
    \end{tabular}%
  \label{tab:solar-count}%
\end{table}%

\begin{table}[h]
  \centering
  \scriptsize
  \caption{Solar roof result by Area.}
    \begin{tabular}{lrrrrrrrr}
    \toprule
    \textbf{Region} & \textbf{Area} & \textbf{Bldg.} & \textbf{Truth} & \textbf{Pred.} & \textbf{Matching} & \textbf{\%Matching} & \textbf{\%FP} & \textbf{\%Cover} \\
    & \textbf{(km$^{2}$)} & \textbf{Area (m$^{2}$)} & \textbf{(m$^{2}$)}& \textbf{(m$^{2}$)}& \textbf{(m$^{2}$)} &  &  & \\
    \midrule
    Berlin 1 & 2.31  & 809753 & 74151 & 113425 & 69314 & 93.48 & 5.45 & 8.56 \\
    Berlin 2 & 1.22  & 669506 & 75217 & 106371 & 72403 & 96.26 & 5.07 & 10.81 \\
    Copenhagen & 1.64  & 465472 & 18885 & 21211 & 15386 & 81.47 & 1.25 & 3.31 \\
    Marseille & 3.51  & 612603 & 63316 & 67446 & 61976 & 97.88 & 0.89 & 10.12 \\
    Melbourne & 4.52  & 1042319 & 76263 & 184071 & 76263 & 100.00 & 10.34 & 7.32 \\
    New York 1 & 1.54  & 449264 & 41442 & 41442 & 41442 & 100.00 & 0.00 & 9.22 \\
    New York 2 & 5.52  & 1746588 & 136615 & 123048 & 113901 & 83.37 & 0.52 & 6.52 \\
    D.C. & 2.10  & 905803 & 161260 & 268043 & 161260 & 100.00 & 11.79 & 17.80 \\
    \midrule
    \textbf{Average} &       &       &       &       &       & \textbf{96.25} & \textbf{4.71} & \textbf{10.20} \\
    \bottomrule
    \end{tabular}%
  \label{tab:solar-area}%
\end{table}%

\begin{figure}[h]
\centering\includegraphics[width=\linewidth]{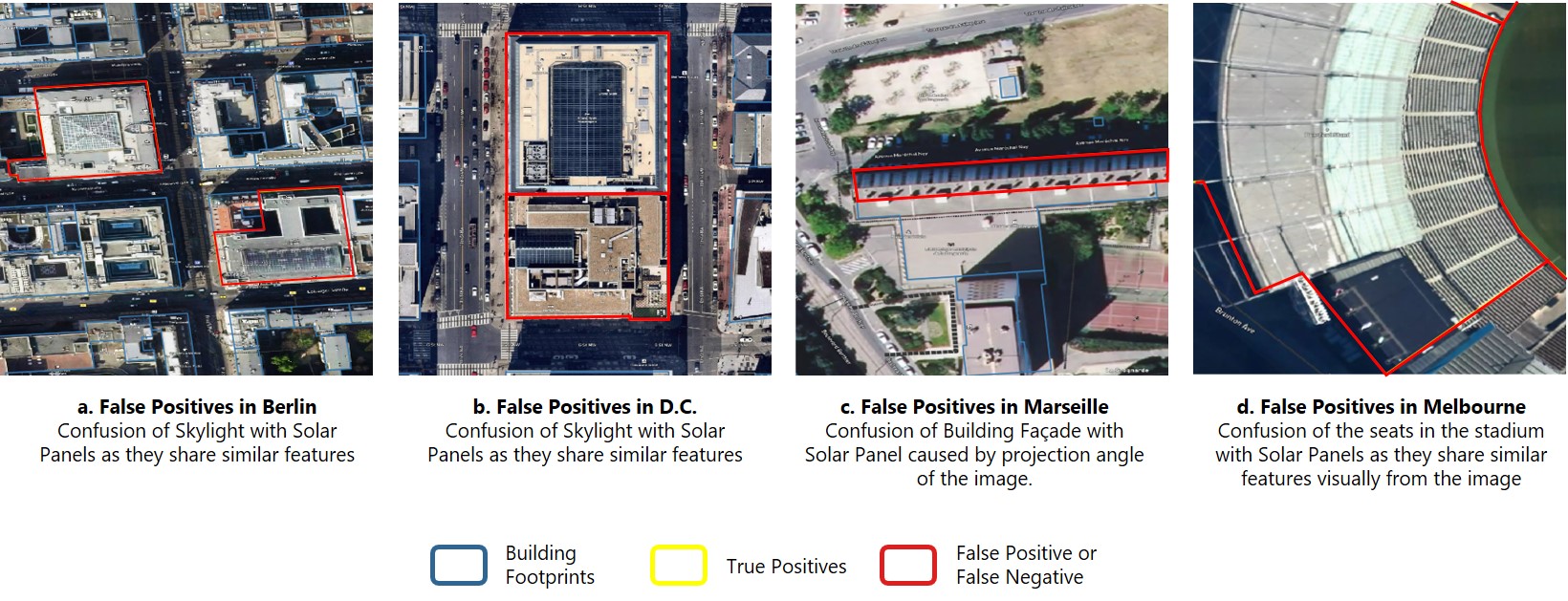}
\caption{Examples of solar roof model confusions.}
\label{fig:solar-confusions}
\end{figure}

There is some confusion between building skylights and solar panels as they have almost identical features and difficult to distinguish also by humans. This issue is the major cause of the rise of \%FP in Berlin (8a) and Washington D.C. (8b) as illustrated in Figure~\ref{fig:solar-confusions}. Another factor that causes confusion is the angle of projection of tall buildings. When the perspective of the satellite photo is not ortho-rectified, the image may show the building facade where the glass windows can be confused with solar panels as seen in  Marseille (8c). Lastly, objects that share similar features with solar panels such as dark rectangles with white border might also be counted as false positives. These objects could be rooftop sunshades or stadium seats such as the example in Melbourne (8d). 

\subsubsection{Evaluation conclusion}

The evaluation results suggest that the percentage of false positives for both solar and green roofs are within acceptable levels, and the two models are sufficiently accurate to provide insightful city-wide spatial and quantitative analyses, both in the location and size of sustainable instruments on rooftops.
In future work, it would be worthwhile to expand our approach by integrating other forms of data. For example, point clouds obtained from airborne Light Detection and Ranging (LiDAR) might provide further information that would aid the classification \citep{rs70912135}, such as 3D geometry and intensity of the returns, which may hint at the characteristics of the surface.

\section{Roofpedia registry and index}\label{sc:index}

The results from the predictions on the 17 cities (Section~\ref{sc:mapping}) are used to create (i) a prototype of a sustainable roof registry, i.e.\ open dataset and web viewer (Section~\ref{sc:registry}); and (ii) an index to assess the penetration of sustainable roof typologies in cities (Section~\ref{sc:sri}).

\subsection{Roof registry}\label{sc:registry}

\begin{figure}[h]
     \begin{subfigure}[b]{\textwidth}
         \centering
         \includegraphics[width=\linewidth]{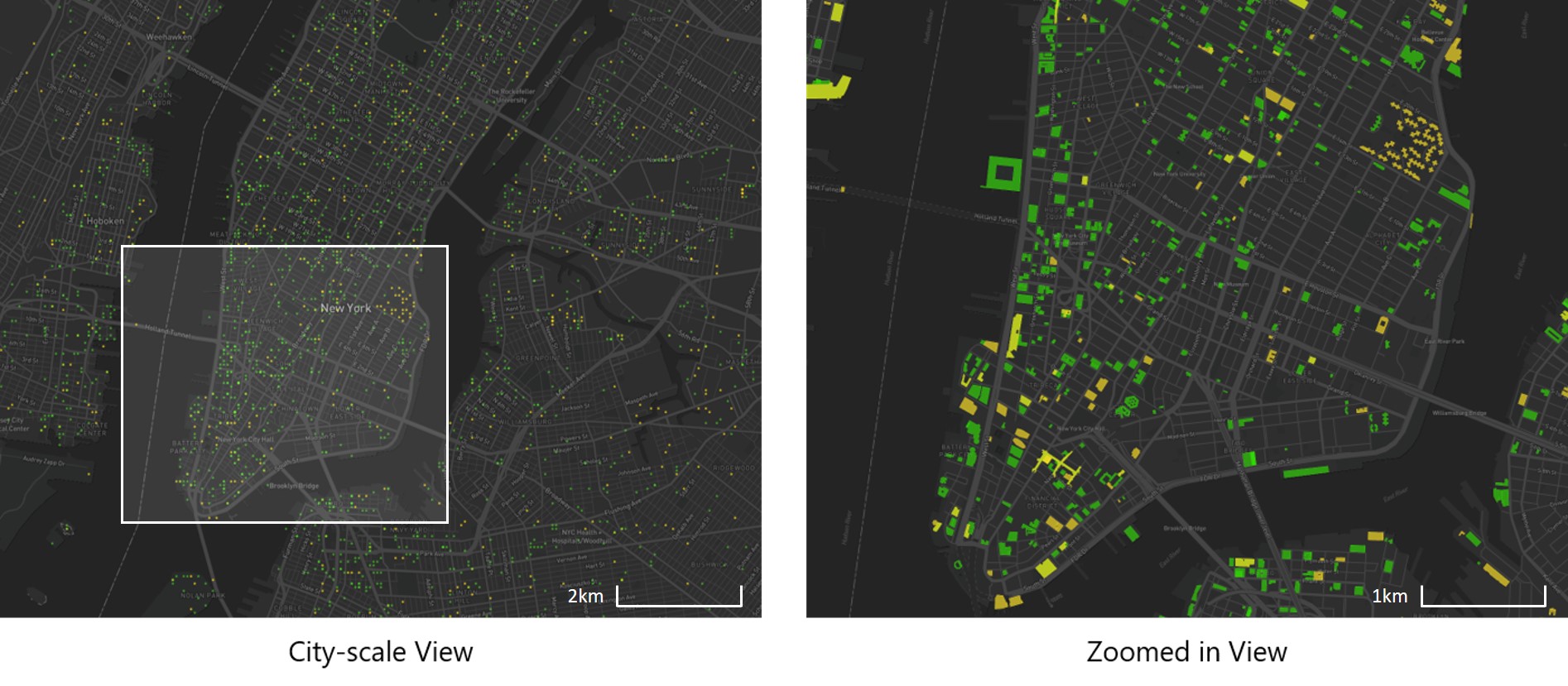}
         \caption{Mapping buildings with sustainable rooftops (example for New York City).}
         \label{fig:nyc-vis}
     \end{subfigure}
     
    \begin{subfigure}[b]{\textwidth}
         \centering
         \includegraphics[width=\linewidth]{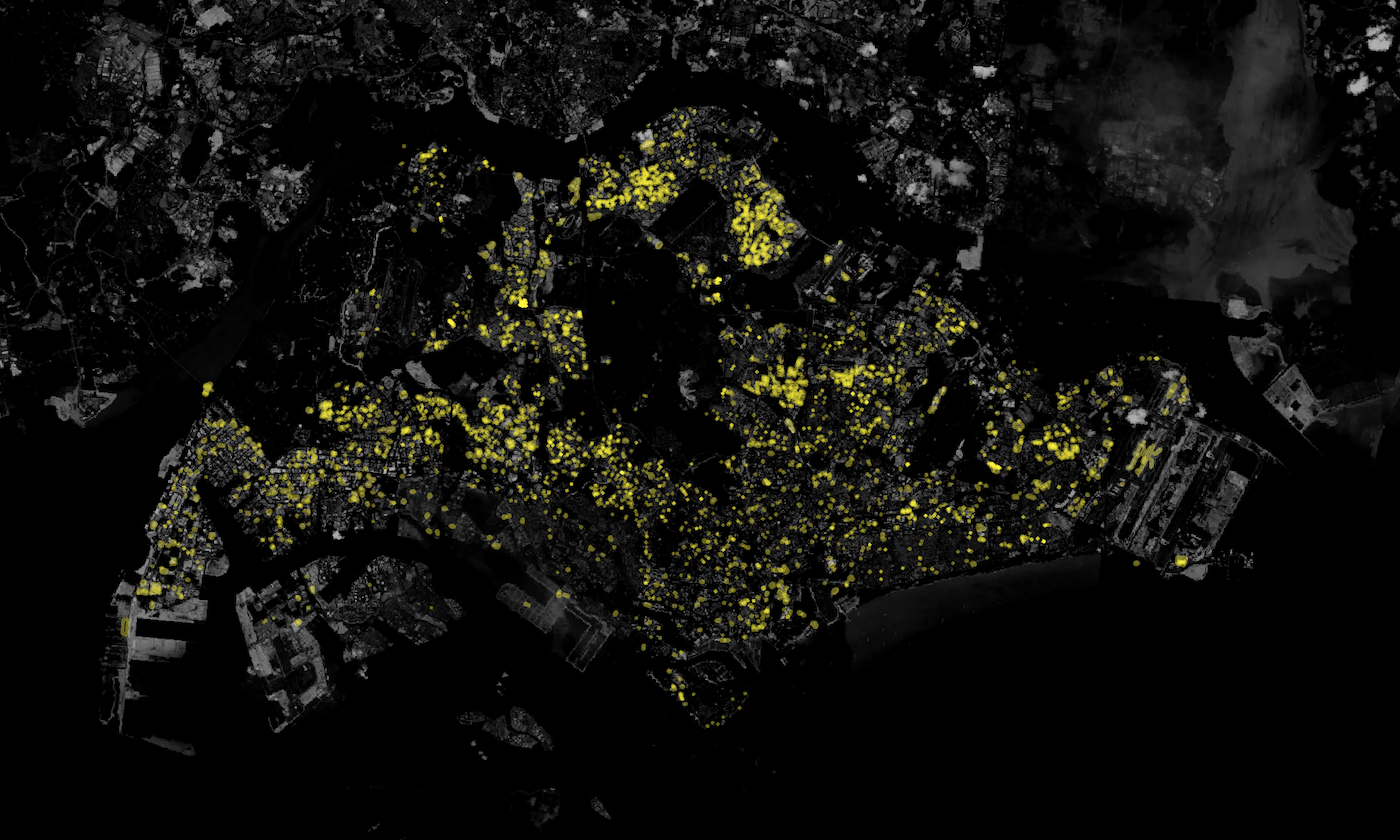}
         \caption{A visualisation for understanding the spatial density of solar roofs (example for Singapore).}
         \label{fig:sg-vis}
     \end{subfigure}
\caption{Visualisation of the spatial distribution of buildings with sustainable rooftops in different styles. Basemap: OpenStreetMap contributors and Mapbox.}
\label{fig:spatial-distribution}
\end{figure}

The roof registry is derived for the 17 cities and it is released as an openly accessible interactive map (Figure~\ref{fig:spatial-distribution}) that visualises the spatial distribution of green and solar roofs in two styles. At smaller scale levels, the centroids of the buildings are shown as yellow (solar roofs) or green (green roofs) dots for easy identification of the city-wide distribution pattern and reveal spatial patterns at the urban scale. At larger scales, the building polygons are shown to highlight buildings with sustainable roofs and addresses can be searched.
Further, the generated data can be visualised as a heatmap to understand the density of sustainable rooftops.

\subsection{Roofpedia Index --- a measure for rooftop sustainability across cities}\label{sc:sri}

The idea of the index is comparable to the one described by the Treepedia project \citep{senseablecitylab}, which measures the street canopy cover in different cities with a Green View Index \citep{li_2015_assessing,seiferling2017green}. In a similar fashion, we create a city-scale summarised index for the roofscape, which assigns scores based on the density and area coverage of green and solar roofs in a city. We hope to raise curiosity and awareness in rooftop utilisation among the public and also provide planners and researchers with a barometer to access the effectiveness of roofscaping policies in different cities.
As indicated in the previous section, the percentage cover of green and solar roofs are calculated based on Count and Area separately. Calculating the percentage by Count helps us to understand the degree of adoption by individual owners while calculating the percentage by Area indicates the extensiveness of the cover. 
For example, the evaluation subset in New York City (NYC2) has a low coverage of green roofs in terms of Count (3.51\%) but a much higher coverage in terms of Area (27.59\%). Such discrepancy suggests that green roofs in NYC are mostly installed on several large buildings while smaller buildings have yet to adopt green roofs. Conversely, although the evaluated subsets Marseille and Berlin 2 have a similar percentage of the cover by area, Berlin 2 has 2.9 times higher \% by Count (3.72\% vs 1.28\%). This difference suggests that the density of buildings with solar roofs is higher in Berlin and more homeowners and developers have adopted solar panels on their roofs.

The Roofpedia Index recognises that both factors are important and takes them into consideration. \%Count (Equation~\ref{eqn4}) and \%Area (Equation~\ref{eqn5}) for both solar and green roof penetration are calculated for each city and normalised with Equation~\ref{eqn6} and Equation~\ref{eqn7}. A combined score is calculated for both solar and green roofs with Equation~\ref{eqn8} and Equation~\ref{eqn9}. An overall score that aggregates both the Solar Score and Green Score is then calculated by taking the average of the sum of two scores as in Equation~\ref{eqn10}.

\begin{align}
\label{eqn4}
& \textrm{\%Count} = 100 \times (\textrm{Matching Count (TP)}/\textrm{Total Count})\\
\label{eqn5}
& \textrm{\%Area} = 100 \times (\textrm{Matching Area (TP)}/\textrm{Total Area})\\
\label{eqn6}
& \textrm{Score by Count} = 100 \times  (\textrm{\%Count} - \textrm{min(\%Count)})/(\textrm{max(\%Count)} - \textrm{min(\%Count)})\\
\label{eqn7}
& \textrm{Score by Area} =  100 \times (\textrm{\%Area} - \textrm{min(\%Area)})/(\textrm{max(\%Area)} - \textrm{min(\%Area)})\\
\label{eqn8}
& \textrm{Solar Score} = (\textrm{Score by Count}_{\textrm{Solar}} + \textrm{Score by Area}_{\textrm{Solar}})/2\\
\label{eqn9}
& \textrm{Green Score} = (\textrm{Score by Count}_{\textrm{Green}} + \textrm{Score by Area}_{\textrm{Green}})/2\\
\label{eqn10}
& \textrm{Overall Score} = (\textrm{Solar Score} + \textrm{Green Score})/2
\end{align}

Table~\ref{tab:green-index} and Table~\ref{tab:solar-index} show the ranking of Solar and Green Roof coverage of the 17 cities and Table~\ref{tab:overall-index} shows the combined ranking of both scores, deriving a holistic index indicating the proliferation of sustainable roofscapes in cities.  

\begin{table}[tbp]
  \centering
  \scriptsize
  \caption{Green Roof Index.}
    \begin{tabular}{rlrrrrrrr}
    \toprule
    \textbf{Rank} & \textbf{City} & \textbf{Bldg.} & \textbf{Green} & \textbf{\%Count} & \textbf{\%Area} & \textbf{Score by} & \textbf{Score by} & \textbf{Green} \\
    & & & \textbf{Roofs}  & & & \textbf{Count} & \textbf{Area} & \textbf{Score}\\
    \midrule
    1     & Zurich & 18440 & 5760  & 31.2  & 41.6  & 100   & 99    & 100 \\
    2     & Berlin & 28677 & 3899  & 13.6  & 24.8  & 43    & 58    & 51 \\
    3     & New York & 34385 & 1924  & 5.6   & 17.2  & 17    & 39    & 28 \\
    4     & Copenhagen & 15505 & 735   & 4.7   & 13.1  & 14    & 29    & 22 \\
    5     & Paris & 74014 & 2766  & 3.7   & 11.2  & 11    & 25    & 18 \\
    6     & San Diego & 28303 & 373   & 1.3   & 11    & 4     & 24    & 14 \\
    7     & San Jose & 182314 & 2650  & 1.5   & 10.2  & 4     & 22    & 13 \\
    8     & Phoenix & 15217 & 245   & 1.6   & 9.9   & 4     & 21    & 13 \\
    9     & Melbourne & 16809 & 258   & 1.5   & 8.4   & 4     & 18    & 11 \\
    10    & Las Vegas & 20389 & 192   & 0.9   & 7.8   & 2     & 16    & 9 \\
    11    & Seattle & 81044 & 347   & 0.4   & 5.6   & 1     & 11    & 6 \\
    12    & Los Angeles & 50978 & 419   & 0.8   & 4.7   & 2     & 9     & 6 \\
    13    & Luxembourg City & 11131 & 125   & 1.1   & 2.7   & 3     & 4     & 4 \\
    14    & Portland & 122900 & 302   & 0.2   & 3.7   & 0     & 6     & 3 \\
    15    & San Francisco & 165814 & 389   & 0.2   & 2.6   & 0     & 3     & 2 \\
    16    & Vancouver & 163818 & 108   & 0.1   & 1     & 0     & 0     & 0 \\
    \bottomrule
    \end{tabular}%
  \label{tab:green-index}%
\end{table}%

\begin{table}[tbp]
  \centering
  \scriptsize
  \caption{Solar Roof Index.}
    \begin{tabular}{rlrrrrrrr}
    \toprule
    \textbf{Rank} &\textbf{City} & \textbf{Bldg.} & \textbf{Solar} & \textbf{\%Count} & \textbf{\%Area} & \textbf{Score by} & \textbf{Score by} & \textbf{Solar} \\
    & & & \textbf{Roofs}  & & & \textbf{Count} & \textbf{Area} & \textbf{Score}\\
    \midrule
    1     & Las Vegas & 20389 & 805   & 3.9   & 17.3  & 86    & 85    & 86 \\
    2     & Zurich & 18440 & 838   & 4.5   & 12.9  & 100   & 61    & 81 \\
    3     & Singapore & 51750 & 1222  & 2.4   & 20.0  & 51    & 99    & 75 \\
    4     & Phoenix & 15217 & 576   & 3.8   & 14.1  & 82    & 67    & 75 \\
    5     & Melbourne & 16809 & 486   & 2.9   & 17.3  & 62    & 85    & 74 \\
    6     & Berlin & 28677 & 809   & 2.8   & 11.3  & 61    & 52    & 57 \\
    7     & Copenhagen & 15505 & 354   & 2.3   & 9.0   & 49    & 40    & 45 \\
    8     & New York & 34385 & 677   & 2.0   & 9.4   & 42    & 42    & 42 \\
    9     & Paris & 74014 & 1507  & 2.0   & 9.1   & 43    & 40    & 42 \\
    10    & San Diego & 28303 & 237   & 0.8   & 7.4   & 16    & 31    & 24 \\
    11    & Los Angeles & 50978 & 384   & 0.8   & 6.3   & 14    & 25    & 20 \\
    12    & Seattle & 81044 & 263   & 0.3   & 5.4   & 5     & 20    & 13 \\
    13    & San Jose & 182314 & 732   & 0.4   & 4.9   & 7     & 17    & 12 \\
    14    & Portland & 122900 & 482   & 0.4   & 4.2   & 6     & 14    & 10 \\
    15    & San Francisco    & 165814 & 560   & 0.3   & 3.9   & 5     & 12    & 9 \\
    16    & Luxembourg City & 11131 & 73    & 0.7   & 1.9   & 12    & 1     & 7 \\
    17    & Vancouver & 163818 & 145   & 0.1   & 1.6   & 0     & 0     & 0 \\
    
    \bottomrule
    \end{tabular}%
  \label{tab:solar-index}%
\end{table}%

\begin{table}[tbp]
  \centering
  \scriptsize
  \caption{Roofpedia Index (overall ranking).}
    \begin{tabular}{rlrrr}
    \toprule
    \textbf{Rank} & \textbf{City} & \textbf{Solar Score} & \textbf{Green Score} & \textbf{Overall Score} \\
    \midrule
    1     & Zurich & 81    & 100   & 91 \\
    2     & Berlin & 57    & 51    & 54 \\
    3     & Las Vegas & 86    & 9     & 48 \\
    4     & Phoenix & 75    & 13    & 44 \\
    5     & Melbourne & 74    & 11    & 43 \\
    6     & New York & 42    & 28    & 35 \\
    7     & Copenhagen & 45    & 22    & 34 \\
    8     & Paris & 42    & 18    & 30 \\
    9     & San Diego & 24    & 14    & 19 \\
    10    & Los Angeles & 20    & 6     & 13 \\
    11    & San Jose & 12    & 13    & 13 \\
    12    & Seattle & 13    & 6     & 10 \\
    13    & Portland & 10    & 3     & 7 \\
    14    & Luxembourg City & 7     & 4     & 6 \\
    15    & San Francisco & 9     & 2     & 6 \\
    16    & Vancouver & 0     & 0     & 0 \\
    \bottomrule
    \end{tabular}%
  \label{tab:overall-index}%
\end{table}%

The results offer material for discussion and further deliberations. 
Looking at the ranking for green roofs, Zurich tops the chart with 41.6\% cover in Area and 31.2\% cover in Count. This impressive lead in green roof coverage affirms the efforts taken by the Zurich City Government in making Green Roofs mandatory for all new buildings since 1991 \citep{tiefbauundentsorgungsdepartement_dachbegrnung}.
Berlin comes in second place with approximately 24.8\% in Area and 13.6\% in Count. This result also presents a substantial lead compared to the rest of the cities in the study and echoes the long tradition of green roofs in Berlin as early as the beginning of the 19th century \citep{ahrendt_2007_historische}.
New York ranks highest in the U.S. with the largest green roof cover. With the new Climate Mobilization Act \citep{nycmayorsofficeofsustainability_2019_legislation}, which made solar or green roofs mandatory on all new construction as well as buildings undertaking major roof renovations in the city, it is highly likely that New York will continue to maintain its edge in a sustainable roofscape and increase its score in the years to come.

In the case of solar roofs, Las Vegas tops the chart with 17.3\% coverage in Area. This result is probably due to the high solar potential of the geography. Phoenix is also another city in the U.S. that leverages its high solar potential throughout the year to harness solar energy.
Other highly ranked cities in the study offer incentives to subsidise solar panel installation, maintaining our vision that this work can be used to assess and monitor the effectiveness of policies. For example, homeowners in Melbourne can enjoy up to AUD 1850 plus the option of an interest-free loan for their solar roof installation until mid-2021 under the Solar Victoria incentive \citep{solarvictoria_2020_solar}, and residents in Copenhagen enjoy tax deductions of up to DKK 7000 to encourage the installation of renewable energy plants such as solar panels \citep{bundgaard_2011_denmarks}.

It is important to note that the Roofpedia Index is not created to promote competition among cities as each city has its unique characteristics, and the exact benefit of greenery or solar panels on rooftops much depends on the urban morphology~\citep{Ng.2012}.
The adoption of these rooftop typologies is also affected by the geolocation and macroclimate of the city. In drier areas, green roofs are harder to maintain while in rainy and dark areas, solar roofs might not make economic sense. 
Taking these limitations into consideration, a city could still be environmentally progressive without a sustainable roofscape. For example, although Vancouver did not perform well in our index, it is nevertheless consistently ranked as one of the most sustainable cities in the world. 
According to the Sustainable Cities Index 2016 \citep{arcadis}, Vancouver ranks highest among North American cities in terms of environmental sustainability. Further, according to Treepedia, Vancouver is one of the cities highest on the list in terms of green canopy. Finally, Vancouver has access to plenty of hydropower, which provides 25\% of the city's energy need alone, and has plans to derive 100\% of the energy used from renewable sources before 2050 \citep{vancouver_2015}. 
Therefore, we believe that the Roofpedia Index complements existing sustainability indices (e.g.\ \citep{Phillis.2017}) by adding a new dimension of consideration in assessing the overall sustainability of a city.

\section{Discussion and limitations}\label{sc:discussion}

We have demonstrated the feasibility and reliability of a scalable method to identify sustainable roofscapes using computer vision and geospatial techniques. The result from the 17 cities has also revealed insights into the distribution and penetration of sustainable roofscapes in these urban areas. 
Besides the perennial importance of space utilisation in the built environment and the positive role of greenery and solar panels, our work is also important at a time of the increased collocation of photovoltaic and farming systems, and the growing importance of urban agriculture \citep{Tablada.2018,Ciriminna.2019,Langemeyer.2021}. 

We have shown that a Convolutional Neural Network could be an alternative approach in roof typology detection apart from traditional classification methods on satellite images. The advantage over the traditional methods is that CNNs are more forgiving of image quality. Where traditional classification methods require multi-spectral satellite imagery that are only available for select cities, our method can cover more cities with RGB images that are available at a reasonable degree of resolution. As such data is becoming available for more locations, additional cities can be added to the registry and index. 
Our modular pipeline does not only allow new cities to be added, but also new roof typologies. Adding such requires training custom models, which can be plugged in the system to predict an additional aspect of the content of roofs.

However, one challenge we have encountered in the process is the impact of inconsistent image quality across cities on prediction accuracy. Running the prediction pipeline on cities with lower quality images results in excessive noise, rendering the prediction unreliable. For example, when predicting rooftop greenery in Singapore, the targets were too blurry even for humans to discern. Fortunately, due to their darker colours and more defined edges, solar panels were much more perceptible than green roofs, thus, we have included Singapore in the Solar Index and the registry of solar rooftops. A general rule of thumb to follow is that as long as a human could discern the rooftop typology effortlessly and create labels without inferring from the surrounding context, the model would be able to automate the detection accurately. The issue of human legibility is also present in the labelling of datasets. When labelling solar panels on rooftops, we are unable to discern whether the solar panel belongs to a solar thermal module. As such, there is yet an effective way to separate solar thermal modules from generic solar panels as they are hard to differentiate even by humans. 

Another consideration in interpreting the result of the predictions is the correlation between the percentage of cover area and the actual roof area covered by greenery and solar panels. In our method, the entire footprint area of a building is counted as the cover area because we assumed that the size of the roof area is proportional to the size of rooftop greenery and solar panels. While this is largely true, there are cases where the actual coverage of rooftop greenery and solar panels only covers a small proportion of the roof area.

On the note of building footprints, it is important to keep in mind that they are sourced from OpenStreetMap, a volunteered geospatial dataset with global coverage. While the quality is sufficiently good, the crowdsourcing nature and myriads of contributors across the world may have different approaches in mapping buildings. For example, a set of adjacent buildings (e.g.\ terraced houses) might be mapped as a single building by one mapper, while another one might map them as multiple individual buildings.
These different modelling approaches might have an impact on the results.

While interpreting the prediction results, we did not take false positives into consideration. As shown in Section~\ref{sc:results}, \%FP fluctuates across cities and could potentially reach 11.79\% in terms of Area and 2.49\% in terms of Count. Although the average \%FP is well below 2\% in terms of count and 4\% in terms of Area, a city with a high \%FP could be placed at a higher position in the index mistakenly, affecting interpretation of results. Therefore, quick manual inspections are still required on the predicted dataset to make sure the results are not exceptionally skewed. On a broader scope, this line of research also grants an opportunity for the public to participate in the mapping process. Just like OpenStreetMap, an interactive platform can be created for the public to verify the result of the AI models to eliminate false positives and the models could be then retrained with a larger dataset to improve its accuracy and precision iteratively.

It is important to recognise that this work is focused on rooftops, and it is geared towards elevating the prominence of the roofscape in the sustainability context. However, they are not the only venue for greenery and PV panels. For example, solar panels may be installed on walls of buildings~\citep{Saretta.2020}, but also on other platforms in cities, e.g.\ noise barriers~\citep{Zhong.2020} and vehicles \citep{Centeno_Brito_2021}. The same goes for urban greening efforts, i.e.\ vertical spaces of buildings such as walls~\citep{Tan.2014,Song:2018hg,Palliwal.2021,Huang.2021}. Furthermore, innovations in BIPV technology have created panels that mimic the appearance of traditional building materials~\citep{verberne2014bipv}. Therefore, even if such panels are installed on building rooftops, it is difficult for both human and the model to detect their presence. A possible direction for future work is to complement our work with other forms of urban data that provide another perspective and cover further venues, e.g.\ using street view imagery \citep{Tang.2019,Ding.2021}.

Our work has further benefits for use cases in different domains, predominantly in energy and urban planning. For example, the city-wide dataset that we have generated may aid researchers to focus on studies at the district or urban scale, such as large-scale retrofit studies~\citep{Ang.2020,Wang:2020jg,Johari:2020ky}.
The dataset may also aid urban planners in understanding the spatial distribution of sustainable roof typologies, e.g. associating them with socio-economic attributes of neighbourhoods. Furthermore, it can provide location and density information (Figure~\ref{fig:spatial-distribution}) for further studies on the impact of green and solar roofs on urban microclimate as the positive effect of solar panels on the UHI effect remains debatable \citep{10, barron}.

\section{Conclusion}\label{sc:conclusion}

As urban population continues to grow, cities need effective measures to improve their environment for both the residents and the surrounding ecology. The roofscape has a demonstrated potential in contributing to a city's sustainability, such as doubling its use by greening them and having solar panels installed on them. While most studies so far have been focused on estimating the potential of rooftops, Roofpedia enables us to understand the current and actual status of the rooftops, i.e.\ how many of them are currently occupied with solar panels and/or extensive greenery. Such information advances the state of the art, complements existing studies (including those estimating the potential), and it is useful for several purposes, e.g.\ gauging the efficiency of government policies, tracking and monitoring pledges by businesses, verifying the use of subsidies, estimating the current carbon offsetting capacity of cities and benchmarking them, and determining how much of the potential has already been realised.

By mapping the distribution of sustainable rooftops across cities in the world, we hope to raise public awareness on the importance of the role of rooftops in supporting sustainable development. We expect that interest in rooftops will be gaining more traction in urban analytics, and thus, our work and the generated data may provide a solid basis for follow-up studies. While our work is focused on green roofs and those with solar panels, we believe that it paves the way for future work in large-scale identification of further rooftop typologies using computer vision.

Besides the ideas discussed in Section~\ref{sc:discussion}, further opportunities for future work are many.
For example, our method can be used to study the temporal evolution of sustainable rooftops, as its landscape is increasingly dynamic thanks to decreasing costs of technology and widely available satellite images~\citep{Zhang.2019}.
The presented approach can be used to analyse satellite imagery from different epochs to provide the difference and evolution in green and solar installations over time, adding the temporal aspect to this project. Such research line would be welcomed by different use cases, such as understanding the effect of urban policies.
In fact, our work suggests that cities that rank highly in the index usually have incentives for green or solar roofs in place, indicating that Roofpedia could be a useful instrument for urban planners and policymakers.

Another idea we have in mind is to expand our work with a crowdsourced platform, bringing it closer to the public and enable enriching our dataset with additional information that might not be available from satellite imagery, and with information that may widen the work beyond the sustainability context.
For example, it might be beneficial to expand the work into providing information about social uses of rooftops through a web mapping service for the public who would like to explore the roofscape around them, e.g.\ to find rooftops in their neighbourhood that have public access and a nice view.

Further options include enrichment of building datasets using the instance that we generated. For example, as OpenStreetMap and 3D building models support modelling and integrating features such as solar panels~\citep{Biljecki:2021vy,stowell2020harmonised}, it might be worthwhile exploring whether our project could serve as a source for tagging buildings with green or solar installations and/or modelling their geometry.

\section*{Acknowledgements}

The authors gratefully acknowledge the sources of the input datasets and the discussions with Yoong Shin Chow (NUS Urban Analytics Lab).
The reviewers' insightful and helpful comments improved the quality of the article and are highly appreciated.
This research is part of the project Large-scale 3D Geospatial Data for Urban Analytics, which is supported by the National University of Singapore under the Start-Up Grant R-295-000-171-133.

\bibliographystyle{cas-model2-names}

\bibliography{roof}
\end{document}